\documentclass[useAMS,usegraphicx,usenatbib,float]{mn2e}
\usepackage{color}
\title[Distribution of dust around galaxies]
{Distribution of Dust around Galaxies: An Analytic Model}

\author[S. Masaki \& N. Yoshida]{{Shogo Masaki$^{1}$\thanks{E-mail:
shogo.masaki@nagoya-u.jp}\thanks{JSPS Fellow} and Naoki Yoshida$^{2,3}$}\\
$^{1}$Department of Physics, Nagoya University, Chikusa, Nagoya 464-8602,Japan\\
$^{2}$Kavli Institute for the Physics and Mathematics of the Universe, TODIAS, University of Tokyo, Kashiwa 277-8583, Japan\\
$^{3}$Department of Physics, University of Tokyo, Tokyo 113-0033, Japan}
\begin{document}

\date{Accepted . Received ; in original form }

\pagerange{\pageref{firstpage}--\pageref{lastpage}} \pubyear{2010}

\maketitle

\label{firstpage}
 
\begin{abstract}
We develop an analytic halo model for the distribution of dust around galaxies.
The model results are compared with the observed surface dust density profile 
measured through reddening of background quasars in the Sloan Digital Sky Survey (SDSS) reported by \cite{MSFR}.
We assume that the dust distribution around a galaxy is described by a simple power law,
similarly to the mass distribution,
but with a sharp cut-off at $\alpha R_{\rm vir}$ where $R_{\rm vir}$ is the galaxy's virial radius
and $\alpha$ is a model parameter. 
Our model reproduces the observed dust distribution profile very well 
over a wide range of radial distance of $10 - 10^{4} h^{-1}$kpc. 
For the characteristic galaxy halo mass of $2\times 10^{12} h^{-1}M_{\odot}$
estimated for the SDSS galaxies, 
the best fit model is obtained if $\alpha$ is greater than unity, 
which suggests that dust is distributed to over a few hundred
kilo-parsecs from the galaxies. 
The observed large-scale dust distribution profile is reproduced if we assume 
the total amount of dust is equal to that 
estimated from the integrated stellar evolution over the cosmic time.
\end{abstract}


\section{Introduction}
How matter is distributed around galaxies is one of the fundamental
questions in cosmology.
Gravitational lensing provides a powerful method to map the matter 
distribution at small and large length scales~\citep[for a review, see][]{BS01}.
Recent large galaxy redshift surveys, such as the Sloan Digital Sky Survey \citep[SDSS;][]{York00} 
and COSMOS survey, have allowed us to explore the mean surface density profile of galaxies through 
weak lensing techniques 
\citep[e.g., ][]{Sheldon04,Mandelbaum06,MSFR,Leauthaud11}.

\citet[][hereafter MSFR]{MSFR} measured the mean surface matter density profile of the SDSS main galaxies 
with the mean redshift $\langle z \rangle=0.36$ through gravitational lensing magnification of background quasars (QSOs).
They calculated the cross-correlation between the number density of foreground galaxies and the flux magnification of background QSOs.
The cross-correlation function was then converted to the surface matter density profile $\Sigma_m$ of the lens galaxies 
as a function of the projected distance $R$ from the galactic center.
The derived profile is well approximated as $\Sigma_m\propto R^{-0.8}$ at $10{\rm kpc}\la R\la10{\rm Mpc}$.

MSFR also detected the systematic offset between the five SDSS photometric bands 
in magnification of background QSOs; in shorter wavelength, QSOs appear less magnified.
It is interpreted as reddening due to dust in and around foreground galaxies.
Adopting the small Magellanic cloud type dust model for the sample galaxies, 
they derived the mean surface dust density profile of galaxies $\Sigma_d(R)$ from the galaxy-QSO color cross-correlation function.
The shape of the derived $\Sigma_d$ is very similar to that of $\Sigma_m$ at $10{\rm kpc}\la R\la10{\rm Mpc}$,
suggesting that there is a substantial amount of dust in the galactic halos
(see also Chelouche et al. 2007; McGee \& Balogh 2010).
Theoretical models are needed to properly interpret the observationally
inferred dust distribution.

In this Letter, we develop an analytic model based on the so-called halo approach to study
the distribution of dust around galaxies. 
Earlier in \cite{MFY}, we used cosmological $N$-body simulations to study in detail
the matter distribution around galaxies. There, we showed that the observed surface density profile
can be used to determine the characteristic mass of the sample lens galaxies,
and that the mass distributed beyond the galaxies' virial radii contributes
about half of the global mass density. 
We provide a physical model for the dust distribution in this Letter.
Our model is characterized by two key physical parameters; 
one is the host halo mass of the galaxies and the other is the extent of dust distribution.
The former is determined from the observed matter profile \cite[e.g.,][]{Mandelbaum06,HW08,Leauthaud11,MFY},
while the latter can be inferred from the dust distribution.
How far dust is transported from the galaxies is indeed a highly interesting
question. The observed dust profile is well described by a single power law 
over a wide range of distance of from 10kpc to 10Mpc. 
We show that the profile is decomposed into two parts, 
the so-called one-halo and two-halo terms.
We parametrize the one-halo term such that dust is distributed
to $\alpha R_{\rm vir}$ where $R_{\rm vir}$ is the galaxy's virial radius.
Through model fitting, we determine the extension parameter $\alpha$
to be greater than unity.
We discuss the implication for the dust production and transport mechanism into intergalactic space.

\section{THE MODEL}
\subsection{Halo approach}
We present a simple formulation to calculate the surface dust density profile.
Our model is based on the so-called halo approach \citep{Seljak00,CS02}.
The mean surface density $\Sigma_d(R)$ is divided into two terms: 
\begin{equation}
  \Sigma_d(R)=\Sigma_d^{\rm 1h}(R)+\Sigma_d^{\rm 2h}(R),
\end{equation}
where $R$ is the distance in the projected two-dimensional plane.
The one-halo term $\Sigma_d^{\rm 1h}(R)$ arises from the central halo, and the two-halo term $\Sigma_d^{\rm 2h}(R)$ from the neighbouring halos.

The contribution from an individual galaxy halo with mass $M_h$ 
to the one-halo term $\Sigma_d^{\rm 1h}(R)$ is given by the projection of the halo 
dust density profile $\rho_d(r|M_h)$ along the line-of-sight $\chi$:
\begin{eqnarray}
  \Sigma_d(R|M_h)=\int^\infty_{-^\infty} {\rm d}\chi\, \rho_d(r=\sqrt{\chi^2+R^2}~|M_h).
\end{eqnarray}
The one-halo term is then a number-weighted average of $\Sigma_d(R|M_h)$
\begin{eqnarray}
  \Sigma_d^{\rm 1h}(R)&=&\frac{1}{n_{\rm halo}}\int_{M_{\rm min}}^\infty {\rm d}M_h\frac{{\rm d}n}{{\rm d}M_h}\Sigma_d(R|M_h),\\
  n_{\rm halo}&=&\int_{M_{\rm min}}^\infty {\rm d}M_h\frac{{\rm d}n}{{\rm d}M_h},
\end{eqnarray}
where ${\rm d}n/{\rm d}M_h$ is the halo mass function and $M_{\rm min}$ is the threshold halo mass for the sample galaxies.
The threshold mass corresponds to the typical host halo mass of the observed galaxies.

We calculate the two-halo term power spectrum $P_d^{\rm 2h}(k)$ as follows:
\begin{eqnarray}
  \label{pk}
  P_d^{\rm 2h}(k)&=&P_{\rm lin}(k)\nonumber \\
  &\times&\left[\frac{1}{\bar\rho_d}\int_0^\infty {\rm d}M_h\frac{{\rm d}n}{{\rm d}M_h}M_d(M_h)b(M_h)u_d(k|M_h)\right]\nonumber \\
  &\times&\left[\frac{1}{n_{\rm halo}}\int_{M_{\rm min}}^\infty {\rm d}M_h\frac{{\rm d}n}{{\rm d}M_h}b(M_h)u_d(k|M_h)\right],
\end{eqnarray}
where $\bar\rho_d$ is the mean cosmic dust density, $P_{\rm lin}(k)$ is the linear matter power spectrum, $b(M_h)$ is the halo bias factor, $M_d(M_h)$ is dust mass in 
and around a halo with mass $M_h$, and $u_d(k|M_h)$ is the Fourier transform of the 
density profile $\rho_d$ normalized by its dust mass.
The power spectrum is converted to the two-point correlation function via
\begin{equation}
  \xi_d^{\rm 2h}(r)=\frac{1}{2\pi^2}\int_0^\infty {\rm d}k\, k^2\frac{\sin(kr)}{kr}P_d^{\rm 2h}(k).
\end{equation}
Then we obtain the two-halo term of the mean surface density profile
\begin{eqnarray}
  \Sigma_d^{\rm 2h}(R)&=&\bar\rho_d\int^\infty_{-\infty}{\rm d}\chi\,\xi_d^{\rm 2h}(r=\sqrt{\chi^2+R^2})\nonumber\\
  &=&2\bar\rho_d\int_R^\infty {\rm d}r\frac{r\xi_d^{\rm 2h}}{\sqrt{r^2-R^2}}.
\label{eq:twohalo}
\end{eqnarray}

We adopt a flat-$\Lambda$CDM cosmology, with $\Omega_m=0.272, \Omega_\Lambda=0.728, H_0=70.2{\rm km~s^{-1}~Mpc^{-1}}, n_s=0.961$ \citep{wmap7}.
We use the code {\it CAMB} to obtain the linear matter powerspectrum \citep{CAMB}
and utilize the halo mass function and bias given by \cite{ST99} at $z=0.36$ 
which is equal to the mean redshift of the galaxy sample used in MSFR.

\subsection{Dust distribution profile}
\begin{figure}
  \includegraphics[width=8.5cm]{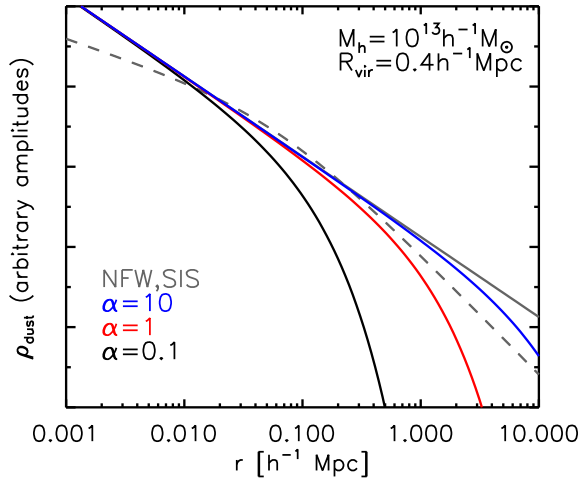}
  \caption{The model dust density profile as a function of the spatial distance from the center. 
The black, red and blue lines represent the model profiles with $\alpha=0.1, 1$ and 10, respectively. 
For comparison, we also show the NFW \citep{NFW} and the untruncated SIS profiles by grey lines. 
Note that the amplitudes are arbitrary in this figure.}
  \label{rho_dust}
\end{figure}

We assume that the spatial distribution of dust within and around a halo 
is organized as 
\begin{eqnarray}
  \rho_d(r|M_h)&\propto&\frac{1}{r^2}\exp\left(-\frac{r}{\alpha R_{\rm vir}}\right).
  \label{eq:dust_1halo}
\end{eqnarray}
where $R_{\rm vir}$ is the virial radius.
Within the virial radius $R_{\rm vir}$, the mean internal matter density is $\Delta\times\rho_{\rm crit}$, 
where $\Delta$ is given by \cite{BN98}.
Essentially, we assume that the dust distribution follows a singular isothermal sphere (SIS) profile 
with exponential cut-off at $r=\alpha R_{\rm vir}$.
One of our aims in this Letter is to determine the value of $\alpha$,
i.e., how far dust is distributed from galaxies.  
Fig.~\ref{rho_dust} shows the shape of the dust density profile $\rho_{\rm d}$ 
for a halo with mass $M_h=10^{13} h^{-1} M_{\odot}$ computed in the above manner.
We see the dependence on $\alpha$ clearly.

The power-law shape is motivated by the fact that the observationally derived
surface dust profile itself is well fit by a simple power law of $\Sigma_d \propto R^{-0.8}$, 
similarly to matter distribution (MSFR). Also, detailed calculations 
of dust ejection and radiation-driven transport by \citet{BF05}
show approximately a power-law distribution for the resulting
gas metallicity through dust sputtering.
We have also examined other profiles of the form $r^{-3}$ and $r^{-1}$
with a similar exponential cut-off. However, we have found that 
neither of the steeper or the shallower profile reproduces well 
the observed dust profile at small distances.
We therefore adopt the profile equation (\ref{eq:dust_1halo}) in our model.

The Fourier transform of $\rho_d(r)$ is given by
\begin{equation}
  u_d(k|M_h)=\int_0^{\infty} {\rm d}r\, 4\pi r^2\frac{\sin(kr)}{kr}\frac{\rho_d(r|M_h)}{M_d(M_h)}.
\end{equation}
Note that the value of $u_d$ should be unity in small-$k$ limit.
We determine the amplitude of $\rho_d$ by setting the 
halo dust mass associated with a halo to be a certain value $M_{\rm d}$. 
To this end, we first consider the total amount of dust around galaxies in the local universe. 
\cite{F11} estimated the total amount, in units of the cosmic density parameter,
\begin{equation}
\Omega_{\rm galaxy~dust}=4.7\times10^{-6}.
\label{eq:globaldust1}
\end{equation}
Interestingly, this value is close to the difference between the estimated
amount of dust produced and shed by stars over the age of the universe 
and the summed amount found in local galactic discs \cite[see also][]{IK03}.
Suppose that the comoving density of the total halo dust remains constant
in the local universe.
Then the mean cosmic density of dust in galactic halos is given by 
\begin{equation}
  \bar\rho_d(z)=\Omega_{\rm galaxy~dust}\, \rho_{\rm crit}(z)(1+z)^3.
\label{eq:globaldust2}
\end{equation}
We set the dust mass associated with a halo to be
\begin{eqnarray}
  \int^\infty_0 {\rm d}r \, 4\pi r^2 \rho_d(r|M_h)=M_d=\Gamma\times M_h
  \label{amp}
\end{eqnarray}
where $\Gamma$ is the dust-halo mass ratio. 
We integrate the dust mass weighted by the halo mass function
to obtain the global dust density. We normalize $\rho_{\rm d}$, 
or equivalently $\Gamma$,
by matching the global dust density to equation (\ref{eq:globaldust2}).
Note that $\Gamma$ is not necessarily a constant 
but can be a function of halo mass. 

\subsubsection{Dust-halo mass ratio $\Gamma$}
An essential physical quantity in our model is 
the dust-halo mass ratio $\Gamma$ in equation (\ref{amp}).
We propose two simple models.
The first one is {\it constant model}, i.e., $\Gamma$ is independent of halo mass.
The dust-halo mass ratio is simply the global density ratio
\begin{equation}
  \Gamma=1.73\times10^{-5}=\Omega_{\rm galaxy~dust}/\Omega_{\rm m}.
\end{equation}

Because the heavy elements that constitute dust are produced by stars,
it may be reasonable to expect that the dust mass 
is proportional to the stellar mass.
Intriguingly, \cite{Takeuchi10} used data of AKARI and GALEX to show 
a moderate correlation between the stellar mass and dust attenuation indicator 
for the sample galaxies (see their Figure 16).
In our second model, we consider the observed galaxy stellar-halo mass relation
to model the halo mass dependence of $\Gamma$.
We call the model as {\it mass dependent model}.
\cite{Leauthaud11} recently studied the stellar-halo mass relation from the joint analysis of 
galaxy-galaxy weak lensing, galaxy clustering and galaxy number densities using the COSMOS survey data.
We use their functional form with the best fit parameters at $z\approx0.37$, 
\begin{eqnarray}
  \log_{10}(M_h)&=&\log_{10}(M_1)+\beta\log_{10}\left(\frac{M_*}{M_{*,0}}\right)\nonumber\\
  &~&+\frac{(M_*/M_{*,0})^\delta}{1+(M_*/M_{*,0})^{-\gamma}}-0.5,
  \label{Mh-Ms}
\end{eqnarray}
where $M_*$ is the galaxy stellar mass, $\log_{10}(M_1/M_\odot)=12.52,\log_{10}(M_{*,0}/M_\odot)=10.92,\beta=0.46,\delta=0.57,$ and $\gamma=1.5$ \citep[see also][]{Behroozi10}.
We then relate the dust mass to the stellar mass as
\begin{equation}
  M_d(M_h) \propto M_*(M_h).
  \label{md-ms}
\end{equation}
The normalization constant is determined to be $3.05\times10^{-3}$ by integrating this equation 
weighted by the halo mass function. The global dust mass density
thus calculated is matched to equation (\ref{eq:globaldust2}).

\begin{figure}
  \includegraphics[width=8.5cm]{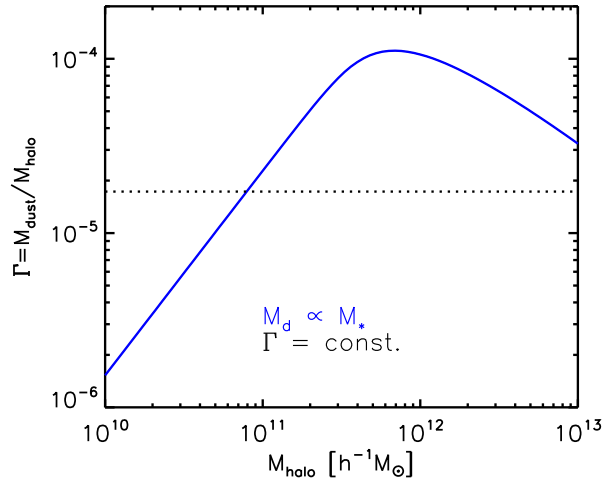}
  \caption{Two models for dust-halo mass ratio as a function of halo mass. 
The dashed and the solid lines are the ratio $\Gamma$ 
for the constant and the mass dependent model, respectively.}
  \label{Gamma}
\end{figure}

Fig. \ref{Gamma} compares $\Gamma$ for our two models.
The shape of $\Gamma$ for the mass-dependent model 
reflects the stellar-halo mass relation. 
The peak value of $\Gamma$ at $\sim 6\times 10^{11} h^{-1}M_{\odot}$ is $\simeq10^{-4}$.
Overall, $\Gamma$ for the mass dependent model is larger 
than that for the constant model at the characteristic
mass of the sample galaxies (see Section 3).

\begin{figure}
  \includegraphics[width=8.5cm]{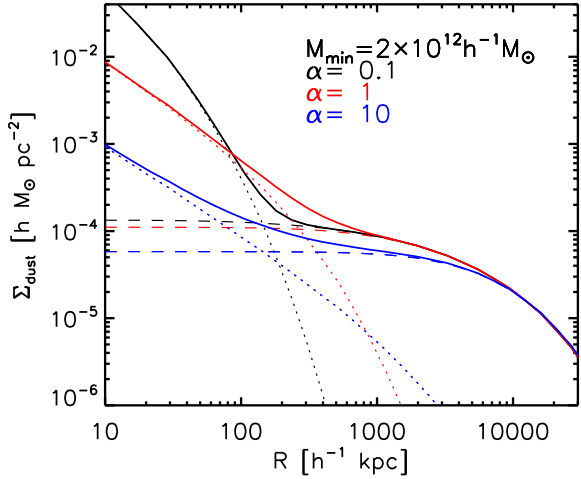}
  \caption{The surface dust density profile as a function of the 
projected radius for $\alpha = 0.1, 1$ and 10. The projected radius is the physical distance 
at $z=0.36$. The constant model for dust-halo mass ratio is adopted. 
The dotted and the dashed lines represent the one-halo and the two-halo terms, respectively. 
The solid lines show the sum of the two terms.}
  \label{sigma_dust_alpha}
\end{figure}

We are now able to calculate the dust surface density profile.
In Fig. \ref{sigma_dust_alpha}, we show the dependence of the 
surface dust density profile on the extension parameter $\alpha$.
For this figure, the threshold halo mass is fixed to be $2\times10^{12}h^{-1}M_\odot$.
We compare three cases; $\alpha=0.1,~1$ and 10.
The dotted lines show the one-halo term.
Clearly the extension parameter $\alpha$ affects significantly 
the amplitude and the shape of the one-halo term.
The central surface density is larger for smaller $\alpha$.
This can be easily understood by noting the total dust mass associated
with a halo is given by equation (\ref{amp}).
On the other hand, $\alpha$ does not affect much the two-halo 
term at $R\ga1{\rm Mpc}$. The amplitude of the two-halo
term is essentially set by the halo bias $b(M_{\rm h})$ (see equation [5]).

\section{RESULTS}
We fit our dust distribution model to the observed surface dust density 
through least chi-squared minimization.
We have two physical parameters, 
$M_{\rm min}$ and $\alpha$, in our model.
We have found that poor constraints are obtained on the parameters 
when both of them are treated as free parameters.
Because $M_{\rm min}$ is already estimated to be $2\times10^{12}h^{-1}M_\odot$ 
in \cite{MFY} through detailed comparison of
the observed matter distribution  
with the results of $N$-body simulations, it is sensible to fit our dust distribution 
model by treating only $\alpha$ as a free parameter. Namely,
the characteristic halo mass can be determined by the gravitational lensing measurement
of the matter distribution, whereas the dust distribution extension can be
inferred from the observed dust profile.

\begin{figure}
  \includegraphics[width=8.5cm]{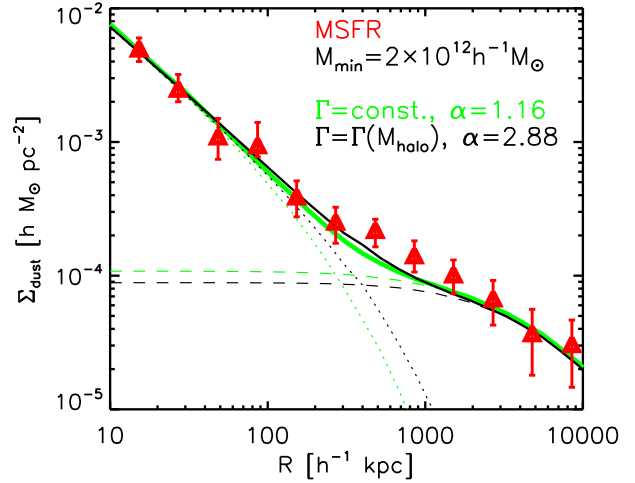}
  \caption{The mean surface dust density profile as a function of the physical projected distance 
from the center of galaxies at $z=0.36$. 
The results from our constant model and mass dependent model 
are shown by the green and the black lines, respectively.
The dotted, the dashed and the solid lines are the one-halo, 
the two-halo and the total, respectively. 
}
  \label{sig_dust}
\end{figure}

We evaluate the likelihood of the specific model by the $\chi^2$ value of the model fit to the observed quantities.
The obtained best-fit $\alpha$ for the constant and the mass dependent models are, respectively,
\begin{eqnarray}
  \alpha&=&1.16^{+0.203}_{-0.155}~(1\sigma)~~{\rm for~constant~model},\\
  \alpha&=&2.88^{+0.450}_{-0.355}~(1\sigma)~~{\rm for~mass~dependent~model}.
\end{eqnarray}
Fig. ~\ref{sig_dust} shows the best-fit dust profile of the constant model with $\alpha=1.16$ 
and that of the mass dependent model with $\alpha=2.88$.
The data points are from MSFR.
Both models for $\Gamma$ reproduce the observed profile fairly well.
It is interesting to compare these two equally good models.
The mass dependent model requires a larger $\alpha$,
which is owing to the difference in the typical value of $\Gamma$
for the two models.
At $M_h > 10^{11}h^{-1}M_\odot$, $\Gamma$ of mass dependent model is higher 
than that of constant model.
Because the one-halo term is largely contributed by halos 
with masses $\sim M_{\rm min} = 2\times 10^{12} h^{-1}M_{\odot}$,
the best fit $\alpha$ is larger for the mass dependent model 
to match the observed inner dust surface density profile (see Fig.\ref{sigma_dust_alpha}).

Overall, our simple models reproduce the observed dust profile very well.
Intriguingly, both our models suggest $\alpha\sim{\cal O}(1)$, i.e., 
halo dust is distributed over a few hundred kilo parsecs 
from the galaxies.
It is also important that the observed power law surface density 
$\Sigma_d\propto R^{-0.8}$
at $R = 10 {\rm kpc} - 10 {\rm Mpc}$ can be explained with 
$\alpha\sim{\cal O}(1)$. 
The apparent large-scale dust distribution
is explained by the two-halo contributions.
Dust is distributed to/over $\sim R_{\rm vir}$ from a galaxy, but not necessarily 
up to 10 Mpc as one might naively expect from the observed dust profile.

It is worth discussing the total dust budget in the universe.
The amplitude of the two-halo term depends largely on the mean 
cosmic density of intergalactic dust, 
$\bar{\rho_{\rm d}}$ in equation (\ref{eq:twohalo})
\footnote{The halo bias $b(M_{\rm h})$ is also 
a critical factor. However, the characteristic halo mass,
and hence $b(M_{\rm h})$, is well constrained from
the observed matter density profile, as shown in \cite{MFY}}.
The excellent agreement at large separation
between the observed dust density profile 
and our model prediction shown in Fig. \ref{sig_dust}
implies that $\Omega_{\rm galaxy~dust}$ should be $\sim10^{-6}$.
Clearly, a significant amount of dust exists around the galaxies.
Such ``halo dust'' can close the cosmic dust budget
as discussed more quantitatively by \citet{F11}.

The intergalactic dust could cause non-negligible extinction 
and thus could compromise cosmological studies with distant supernovae
\citep{Menard10b}.
We calculate the mean extinction by the intergalactic dust 
following \cite{Zu11} (see their equation [2]).
With our model mean cosmic density of $\Omega_{\rm galaxy~dust}=4.7\times10^{-6}$,
the predicted mean extinction is
$\langle A_V \rangle = 0.0090~{\rm mag}$ to $z=0.5$.
Such an opacity is not completely negligible even in the current SNe surveys,
and will become important for future surveys that are aimed at 
determining cosmological parameters with sub-percent precision
\citep{Menard10b}.

\section{Discussion}
We have shown that our halo model can reproduce the dust profile around galaxies measured by MSFR.
By fitting the model to the observed dust profile, 
we infer that dust is distributed beyond the virial radius of a galaxy.
Several authors proposed radiation-driven transport 
of dust from galactic discs into intergalactic medium at high redshifts \citep{Aguirre01,BF05}.
\cite{Zu11} showed that galactic winds can disperse dust into the inter-galactic
medium efficiently. 
Such studies generally suggest that dust can travel up to a few $\times$ 100 kpc from galaxies
if the ejection velocity is $\simeq100{\rm km~s^{-1}}$. 
The relatively larger extent radius of dust for our mass dependent model 
requires very efficient transport mechanisms. 
Note also that the dust must survive on its way through 
the galactic halos. Dust in a large, group-size halo could be
destroyed by thermal sputtering in hot gas (Bianchi \& Ferrara 2005; McGee \& Balogh 2010).
Clearly, detailed theoretical studies on dust transport 
are needed. 

Fluctuations of the cosmic infrared background (CIB) provide insight 
into dust distribution around galaxies \cite[e.g.,][]{Viero09,Amblard11}.
\cite{Viero09} used BLAST data to measure the CIB power spectrum. 
Using a halo approach, they found that the observed power spectrum
at small angular scales is reproduced if halo dust 
extends up to a few times of the virial radius of galactic halos.
It is remarkably consistent with our conclusion in this Letter.
\cite{Amblard11} compared their measurements of the CIB anisotropies 
from Herschel wide-area surveys with \cite{Viero09}.
Two power spectra are consistent with each other at small scales.
Our dust distribution model may provide a key element for studies
on the CIB.

Although our model reproduces the observed dust profiles very well, 
a few improvements can be certainly made.
The dust extension $\alpha$ and the dust-halo mass ratio 
$\Gamma$ are likely to depend on the halo mass and galaxy type etc 
(McGee \& Balogh 2010) .
One may need to consider the distribution of satellite galaxies 
within a halo by using, for example,
the halo occupation distribution (HOD).
Indeed, we see slight discrepancies between the model predictions and the observation 
in the dust profiles at $\sim1$Mpc (Fig. ~\ref{sig_dust}), where the contribution from satellite 
galaxies are non-negligible \citep[for more detailed modeling, see e.g.,][]{Leauthaud11}.
In principle, the HOD parameters can be inferred
from the lensing magnification measurement presented by MSFR.
However, in order to derive the parameters accurately,
one needs to use additional information from observations
of the galaxy-galaxy correlation function \citep{Leauthaud11}. 
Including these effect in our model 
is beyond the scope of this Letter, but will be needed in order to
interpret data from future wide-field galaxy surveys.

\section*{ACKNOWLEDGMENTS}
We thank R.S. Asano, M. Fukugita, A. Leauthaud, B. M{\'e}nard and T.T. Takeuchi for helpful discussions.
SM is supported by the JSPS Research Fellowship. 
NY acknowledges financial support by the Grant-in-Aid for Young Scientists　
(S 20674003) by JSPS. 
The work is supported in part by KMI and GCOE at Nagoya University,
by WPI Initiative by MEXT, and by Grant-in-Aid for Scientific Research on Priority Areas No. 467.



\bsp

\label{lastpage}

\end{document}